\documentclass[prd,aps,showpacs,floats,letterpaper,floatfix,groupedaddress]{revtex4}
\usepackage{amssymb,amsmath}
\usepackage{graphicx,epsf, epsfig, psfig, amssymb}% Include figure files

\usepackage{dcolumn,epsfig}

\begin{document}

\title{Gravitational diffraction radiation}

\author{Vitor Cardoso}
\email{vcardoso@phy.olemiss.edu} \affiliation{Department of Physics and Astronomy, The University of
Mississippi, University, MS 38677-1848, USA \footnote{Also at Centro de F\'{\i}sica Computacional, Universidade
de Coimbra, P-3004-516 Coimbra, Portugal}}

\author{Marco Cavagli\`a} \email{cavaglia@phy.olemiss.edu}
\affiliation{Department of Physics and Astronomy, The University of Mississippi, University, MS 38677-1848,
USA}

\author{M\'ario Pimenta} \email{pimenta@lip.pt}
\affiliation{
\centerline{\hbox{Departamento de F\'{\i}sica, IST, Av.\ Rovisco Pais, 1049-001 Lisboa, Portugal}}\\
and
\\
\centerline{\hbox{LIP, Av.\ Elias Garcia, 14-1, 1000-149 Lisboa, Portugal}}}

\date{\today}

\begin{abstract}
We show that if the visible universe is a membrane embedded in a higher-dimensional space, particles in uniform
motion radiate gravitational waves because of spacetime lumpiness. This phenomenon is analogous to the
electromagnetic diffraction radiation of a charge moving near to a metallic grating. In the gravitational case,
the role of the metallic grating is played by the inhomogeneities of the extra-dimensional space, such as a
hidden brane. We derive a general formula for gravitational diffraction radiation and apply it to a
higher-dimensional scenario with flat compact extra dimensions. Gravitational diffraction radiation may carry
away a significant portion of the particle's initial energy. This allows to set stringent limits on the scale
of brane perturbations. Physical effects of gravitational diffraction radiation are briefly discussed.
\end{abstract}

\pacs{04.30.-w, 04.50.+h, 11.25.Wx, 46.40.Cd}

\maketitle
%%%%%%%%%%%%%%%%%%%%%%%%%%%%%%%%%%%%%%%%%%%%%%%%%%%%%%%%%%%%%%%%%%%%%%%%%%%%%%%%%%%%%%%%%%%%%%%%%%%%%%%%%%%%%
\section{Introduction}
\label{intro}
%%%%%%%%%%%%%%%%%%%%%%%%%%%%%%%%%%%%%%%%%%%%%%%%%%%%%%%%%%%%%%%%%%%%%%%%%%%%%%%%%%%%%%%%%%%%%%%%%%%%%%%%%%%%%
Larmor's formula of electromagnetism \cite{Jackson} states that an electric charge in uniform motion does not
radiate. However, there are two ways to have radiation from a charge moving with constant velocity. The first
way is to have a particle moving in a medium with velocity exceeding the phase velocity of light in that
medium. This gives rise to the well-known Vavilov-Cherenkov radiation \cite{Jackson,GinzburgFrank,Tamm}. The
second way is to consider charge motion in inhomogeneous media. Ginzburg and Frank \cite{GinzburgFrank} first
discussed this effect by investigating a particle in uniform motion which crosses a planar interface between
two media with dissimilar refractive index. This kind of radiation is known as transition radiation
\cite{Jackson}. More generally, any motion near finite-size objects also induces radiation, in a process called
diffraction radiation \cite{Bolotovskii}. One of the first experimental verifications on this effect was
provided by Smith and Purcell \cite{SmithPurcell}. The Smith-Purcell experimental set-up consisted of an
electron moving close to the surface of a metal diffraction grating at right angles to the rulings. The theory
of the Smith-Purcell effect has been discussed by several authors \cite{FranciaBerg}.

The aim of this paper is to show the existence of gravitational diffraction radiation (GDR) and discuss its
physical effects. The model under consideration is a braneworld scenario \cite{Maartens:2003tw} such as, for
example, the Randall-Sundrum I model \cite{Randall:1999ee}. Braneworld models have attracted a lot of interest in
recent years,  revolutionizing our view of how our universe may be described. The central idea of braneworld
scenarios is that the visible universe is restricted to a four-dimensional brane inside a higher-dimensional
space, called the bulk. The additional dimensions are taken to be compact and other branes may be moving through
the bulk. Interactions of the visible brane with the bulk and hidden branes introduce effects not seen in standard
physics.

\begin{figure}
\centerline{\includegraphics[width=0.45\textwidth]{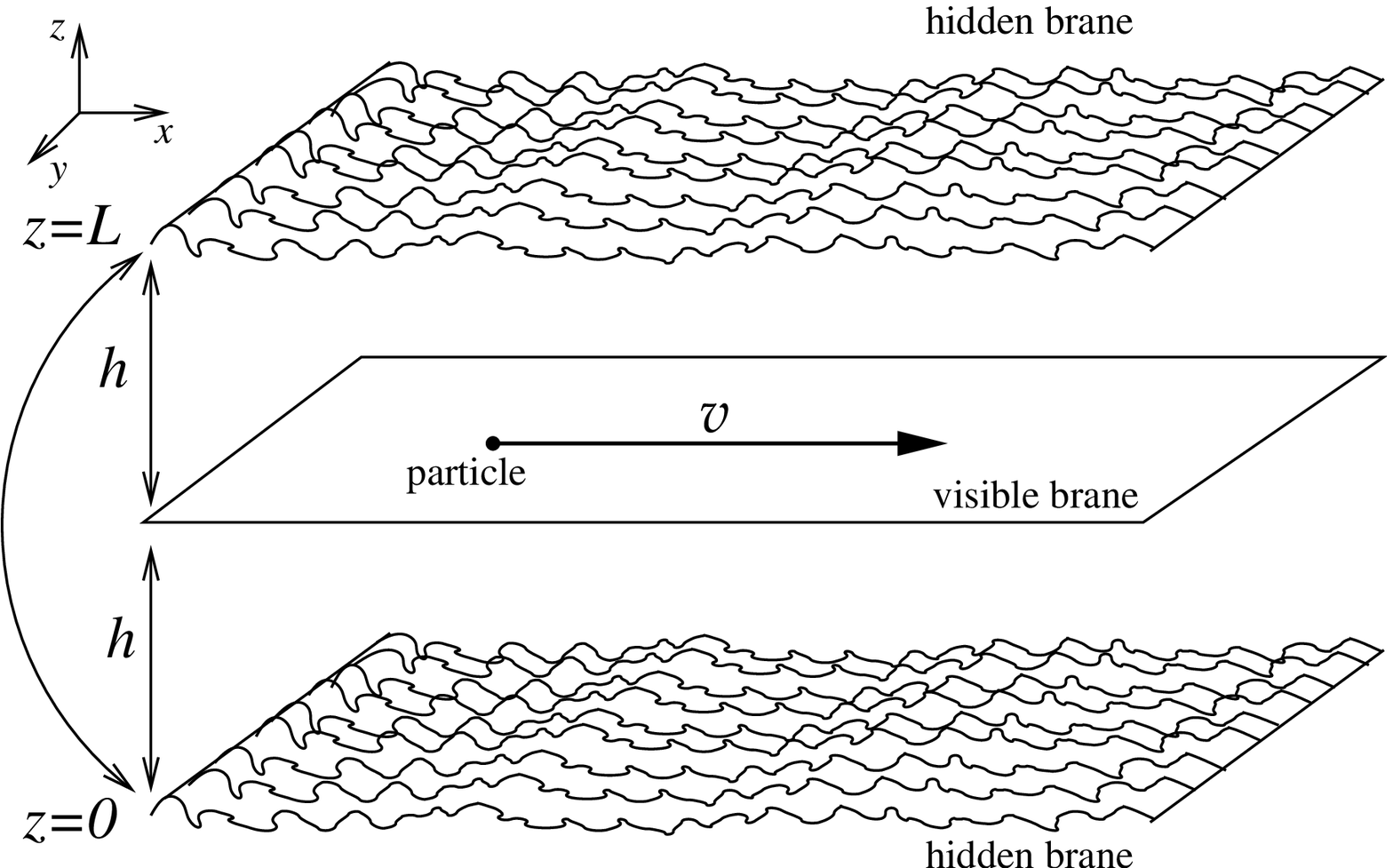}\qquad\qquad
\includegraphics[width=0.30\textwidth]{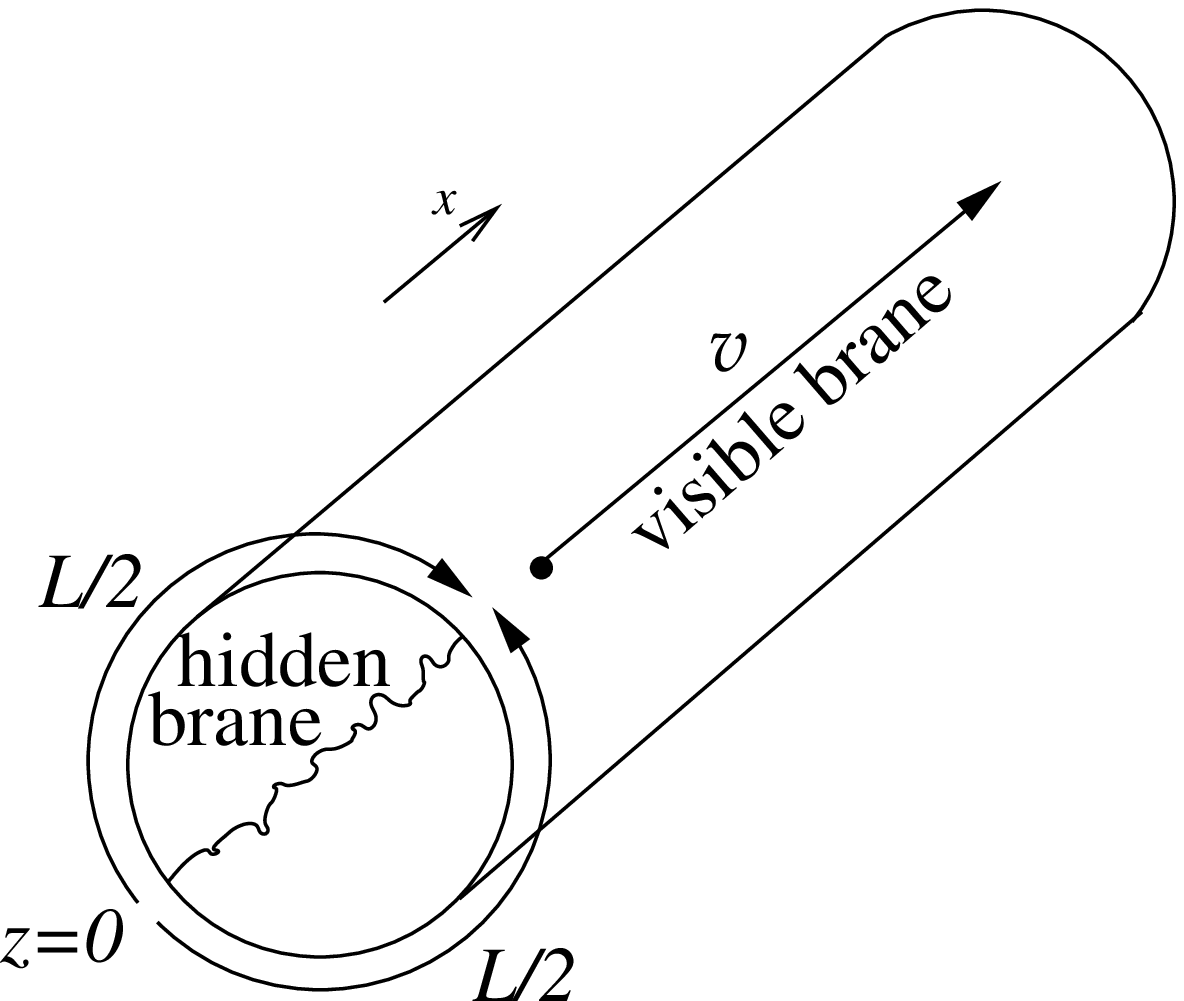}}
\caption{Pictorial representation of the braneworld scenario considered in the text. The hidden brane at $z=0$
is identified with the brane at $z=L$. The visible brane is located at $z=h=L/2$. The particle moves on the
visible brane along $x$ with constant velocity $v$. Only one transverse coordinate, $y$, is shown. The right
panel shows the wrapping around the extra dimension $z$. If the two halves of the cylinder are identified, the
Randall-Sundrum model I \cite{Randall:1999ee} is obtained.}
\label{braneworld1}
\end{figure}

In this set-up, a particle in uniform motion on the visible brane radiates gravitational waves due to the presence
of a second (hidden) brane at finite distance, which plays the role of the metal diffraction grating of the
Smith-Purcell experiment. GDR on the visible brane is generated by inhomogeneities on the hidden brane due, for
example, to bulk-brane interactions and brane fluctuations \cite{Bando:1999di}. We will show that the amount of
GDR depends on the size of extra dimensions and the length scale of brane perturbations. This result is general
and independent on the fine details of the model. Without loss of generality, in our computations we will consider
a flat five-dimensional spacetime with the extra dimension taking values within the interval $[0,L]$. The distance
between the particle and the diffraction grating, $h=L/2$, is the distance between the two branes located at the
orbifold fixed points. The model is illustrated pictorially in Fig.\ \ref{braneworld1}. The generalization to
higher-dimensions is trivial and is discussed at the end of this letter.

%%%%%%%%%%%%%%%%%%%%%%%%%%%%%%%%%%%%%%%%%%%%%%%%%%%%%%%%%%%%%%%%%%%%%%%%%%%%%%%%%%%%%%%%%%%%%%%%%%%%%%%%%%%%%
\section{Theory of Gravitational Diffraction Radiation}
\label{theory}
%%%%%%%%%%%%%%%%%%%%%%%%%%%%%%%%%%%%%%%%%%%%%%%%%%%%%%%%%%%%%%%%%%%%%%%%%%%%%%%%%%%%%%%%%%%%%%%%%%%%%%%%%%%%%
In the harmonic gauge, the five-dimensional linearized Einstein equations are
\begin{equation}
\Box\, h_{\mu\nu}=-16\pi G_5 S_{\mu\nu}\,,
\label{eqmotion1}
\end{equation}
where $h_{\mu\nu}$ represents small corrections to the flat background and $S_{\mu\nu}$ is the modified
energy-momentum tensor of the source \cite{Cardoso:2002pa}. For sake of computational simplicity, we replace the
the gravitational field components with a single scalar degree of freedom, i.e.\ rewrite the above equation as
$\Box\, \varphi=S(x)$. This is a common procedure \cite{Barvinsky:2003jf}. If we carefully
select $S(x)$, the scalar field $\varphi$ can mimic all aspects of gravitational waves (except polarization)
\cite{Misner:1972jf}. The source of the field is a minimally coupled pointlike particle with nonzero mass $m$. If
we represent its worldline by $x^{\mu}=x_p^{\mu}(\tau)$, where $\tau$ is the proper time, $\phi$ satisfies the
inhomogeneous wave equation
\begin{equation}
\Box\, \varphi= g\int d\tau\,
\delta^{5}(x^{\mu}-x_p^{\mu}(\tau))\,,
\label{eqmotion2}
\end{equation}
where $g$ is the coupling constant. We consider a particle in uniform motion on the visible brane. Denoting
with $z$ the coordinate transverse to the brane and with $y_1,y_2$ the longitudinal coordinates perpendicular
to the particle direction of motion $x$, the particle worldline in the brane reference frame is
$x^\mu(\tau)=(t=\gamma\tau,x=vt,y_1=0,y_2=0,z=h)$, where $\gamma$ is the Lorentz factor of the particle. The
time-averaged energy radiated per unit time in the direction $\bf n$ is
\begin{equation}
\frac{dE_{\bf n}}{dt}=-\frac{1}{2}\hbox{Re}\,\left[\int dA\,\dot\varphi\,
\frac{\partial\varphi^\star}{\partial n}\right]\,,
\label{powern}
\end{equation}
where $dA$ is the surface element with normal $\bf n$. If the tensor structure of the gravitational
perturbations is taken into consideration, the source term in Eq.\ (\ref{eqmotion2}) must be changed into
\begin{equation}
S^{\mu \nu} \sim m \int d\tau\, \dot{x}^{\mu} \dot{x}^{\nu} \delta^{5}(x^{\mu}-x_p^{\mu}(\tau))\,.
\label{Smunu}
\end{equation}
The results for the scalar field can be translated to the gravitational case by setting $g=\sqrt{G_5}m\gamma^2$
\cite{Misner:1972jf}.

If the branes are smooth, a particle with constant velocity does not radiate. This can be checked by deriving
the Larmor's formula for the field. In order to avoid complications due to the higher-dimensional nature of the
spacetime \cite{Cardoso:2002pa,Barvinsky:2003jf,SoodakTiersten}, we temporarily assume that there is only one
transverse spatial dimension. The total power emitted per unit of solid angle in the direction $\bf n$ is
\begin{equation}
\frac{dP}{d\Omega}=\frac{g^2}{16\pi^2}\frac{\left[\dot{\bf v}\cdot\left(\gamma^2(1-{\bf v}\cdot{\bf n}){\bf v}-{\bf n}
\right)\right]^2}{\gamma^2 (1-{\bf v}\cdot{\bf n})^3}{\bf n}\,.
\label{powerangle}
\end{equation}
Thus a particle in uniform motion in empty space (on the brane) does not radiate. If a hidden brane is present,
the Green's function has to be modified to take into account its effects. The hidden brane can be thought as a
wall parallel to the particle direction of propagation at distance $h$. By repeating the steps leading to Eq.\
(\ref{powerangle}) it is straightforward to show that there is no radiation for uniform motion parallel to the
wall. This result can be understood by boosting the solution to the rest frame of the particle. The problem is
reduced to a static problem with one image particle on the other side of the wall. Clearly, the reduction to
the static configuration is only possible if the brane is smooth and infinite in the $x$ direction. If the
brane is inhomogeneous in the particle direction of propagation, diffraction radiation is generated. In that
case, the image configuration is time dependent and the system is equivalent to a set of oscillating charges in
the particle reference frame. Diffraction radiation can be understood as being generated by the reflection of
the boosted static field on the nearby wall.

\begin{figure}
\centerline{\includegraphics[width=0.50\textwidth]{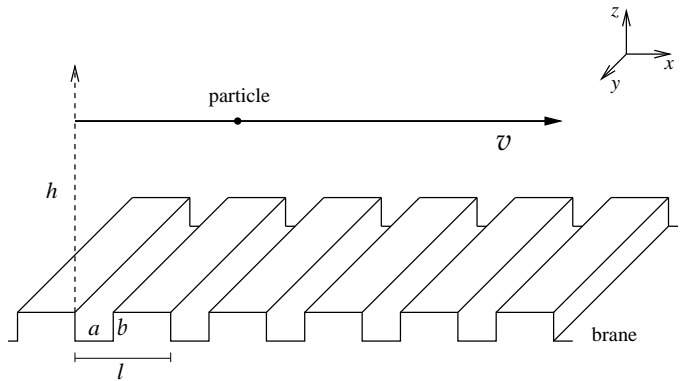}} \caption{Pictorial representation of brane
perturbations. The longitudinal brane perturbations are periodic in $x$ with length $l$. The transverse
perturbations have length $b$. Only one transverse coordinate, $y$, is shown.} \label{braneworld3}
\end{figure}

We now compute the radiated power in the presence of a brane with typical longitudinal perturbations of length
scale $l$ and transverse perturbations of length scale $b$. These perturbations are modeled with a $l$-periodic
lamellar grating with rulings of width $a$ perpendicular to the particle direction of motion as in Fig.\
\ref{braneworld3}. (The Smith-Purcell effect for a single grating with these characteristics has been discussed
in  Ref.\ \cite{Shibata}.) Although this model is clearly an oversimplification, its main features do not
depend on the choice of the perturbation structure: The existence of diffraction radiation is due to the
excitation of propagating modes by evanescent waves and is independent of the particular mechanism by which the
propagating modes are excited.
The calculation with a perturbed brane can be conveniently solved by considering the Fourier transform of the
field $\varphi$:
\begin{equation}
\varphi=\frac{1}{(2\pi)^3}\int d\omega\, d^2\eta\, e^{-i\omega t+i\boldsymbol{\eta}\cdot{\bf y}}
\phi(x,z;\omega,\boldsymbol{\eta})\,.
\label{Fourier}
\end{equation}
The field equation in the Fourier space is
\begin{equation}
\left[\boldsymbol{\nabla}^2+(\omega^2-\boldsymbol{\eta}^2)\right]\phi=g/(\gamma v)\,e^{i\alpha_0
x}\delta(z-h)\,,
\label{Fourier2}
\end{equation}
where $\alpha_0=\omega/v$. The general solution for the field $\phi$ in the bulk is the sum of
the inhomogeneous solution of the above equation plus a superposition of plane wave solutions of the
two-dimensional Helmholtz equation:
\begin{equation}
\phi_{\rm bulk}=A\left\{e^{i\alpha_0 x+i\gamma_0|z-h|}+
\sum_{n=-\infty}^{+\infty}\left[B_n e^{i\gamma_nz}+C_n e^{-i\gamma_nz}\right]e^{i\alpha_n x}\right\}\,,
\label{phi}
\end{equation}
where $A=-ig/(2\gamma\gamma_0 v)$, $\gamma_n=(\omega^2-\eta^2-\alpha_n^2)^{1/2}$ and $\alpha_n=\alpha_0+2\pi
n/l$. Since the parameter $\gamma_0=i[\omega^2/(v\gamma)^2+\boldsymbol{\eta}^2]^{1/2}$ is pure imaginary, the
first term in Eq.\ (\ref{phi}) describes an evanescent wave that decays exponentially with increasing distance
from the particle trajectory. This evanescent wave is the non-propagating part of the spectrum due to the
particle in uniform motion. The second term in Eq.\ (\ref{phi}) represents a superposition of propagating plane
waves. The physical interpretation of these propagating modes is that of radiation arising from the modes of
the brane perturbations, which are excited by the evanescent waves. The amplitude ${\cal A}(\omega)$ of the GDR
is thus expected to be of the order of the amplitude of the non-propagating modes, i.e.,
\begin{equation}
{\cal A}(\omega)\sim
|A|\, e^{i\gamma_0 h}\sim (g/\omega)\,e^{-\omega h/\gamma}\,,
\label{amplitude}
\end{equation}
where we have taken the relativistic limit $\gamma\gg 1$. This can be shown by computing the amplitudes of the
radiating modes as function of the evanescent wave amplitude $A$. The solution in the bulk must match the solution
on the brane
\begin{equation}
\phi_{j,{\rm br}}=A\, \sum_{m=0}^{+\infty}D_{j,m}
\left[e^{-i\kappa_mz}-E_{j,m} e^{i\kappa_mz}\right]\sin\left[\frac{\pi mx_j}{\Delta_j}\right]\,,
\label{phibrane}
\end{equation}
where $\kappa_m=(\omega^2-\eta^2-(\pi m/a)^2)^{1/2}$ and $E_{j,m}=(e^{2i\kappa_m b},e^{-2i\kappa_mL})$,
$x_j=(x,x-a)$ and $\Delta_j=(a,l-a)$ in the intervals $x\in[0,a]$ ($j=1$) and $x\in[a,l]$ ($j=2$),
respectively. Choosing Dirichlet boundary conditions, the amplitudes of the propagating waves satisfy the
system of linear inhomogeneous algebraic equations:
\begin{eqnarray}
&&\sum_{n=-\infty}^{+\infty}\left[B_n(l\delta_{nk}+V_{1,nk})+C_n(l\delta_{nk}-V_{1,nk})\right]+
e^{i\gamma_0L/2}(l\delta_{0k}-V_{1,0k})=0\,,\\
&&\sum_{n=-\infty}^{+\infty}\left[B_ne^{i\gamma_n(L-b)}(l\delta_{nk}+V_{2,nk})+
C_ne^{-i\gamma_n(L-b)}(l\delta_{nk}-V_{2,nk})\right]+ e^{i\gamma_0(L/2-b)}(l\delta_{0k}+V_{2,0k})=0\,,
\label{coeff1}
\end{eqnarray}
where
\begin{equation}
V_{j,nk}=2\Delta_j\epsilon_j\gamma_n\sum_{m=0}^{+\infty}\kappa_m^{-1}\tan(\kappa_m
b)\,\Phi_{j,mk}\Phi^\star_{j,mn}\,,~~~~
\Phi_{j,mk}=\Delta_j^{-1}\int_0^{\Delta_j} dx\,e^{-i\alpha_k x}\,\sin\left(\pi
mx/\Delta_j\right)
\label{coeff2}
\end{equation}
and $\epsilon_j=(1,e^{2i\pi(n-k)a/l})$. The power emitted in the bulk follows from Eq.\ (\ref{powern}) after
solving  Eqs.\ (\ref{coeff1}) for the propagating wave amplitudes $B_n$ and $C_n$. Generalizing to $D$ spacetime
dimensions and taking the relativistic limit, we have
\begin{equation}
P \sim \frac{g^2 b^2}{\gamma^2 l^{D}}\left (1+{\cal O}(b/l)\right ) e^{-2\pi L/(\gamma l)} \,,
\label{fim}
\end{equation}
where we have set $l=2a$ and assumed small $b/l$. Since we assumed the source to be pointlike, Eq.\
(\ref{fim}) is valid for particle size $\ll$ $L$, $b$ and $l$, otherwise it describes the diffraction radiation
of a relativistic particle under very general assumptions. The GDR energy loss per unit distance is
\begin{equation}
\frac{d\ln{E}}{dx}\sim E\frac{G_Db^2}{l^{D}}e^{-2\pi L/(\gamma l)}\,,
\label{dEdx}
\end{equation}
where $G_D$ is the $D$-dimensional Newton constant.
%%%%%%%%%%%%%%%%%%%%%%%%%%%%%%%%%%%%%%%%%%%%%%%%%%%%%%%%%%%%%%%%%%%%%%%%%%%%%%%%%%%%%%%%%%%%%%%%%%%%%%%%%%%%%
\section{Bounds on brane inhomogeneities}
\label{bounds}
%%%%%%%%%%%%%%%%%%%%%%%%%%%%%%%%%%%%%%%%%%%%%%%%%%%%%%%%%%%%%%%%%%%%%%%%%%%%%%%%%%%%%%%%%%%%%%%%%%%%%%%%%%%%%
As an example of the physical consequences of GDR, let us
consider the scenario with $D-4$ large extra dimensions of size $L$ \cite{Arkani-Hamed:1998rs}. In this model
$G_D$ is related to the four-dimensional Newton constant $G_4$ by $G_D\sim L^{D-4}G_4$. There are two natural
length scales, the Planck length, $L_{\rm Pl}\sim 10^{-17}$ cm, and the size of the large extra dimensions,
$L\sim 10^{30/(D-4)-17}$ cm. The perturbation length scales $l$ and $b$ are expected to be larger than the
Planck length. This implies $L/l\lesssim 10^{30/(D-4)}$ and  $b/L\gtrsim 10^{-30/(D-4)}$. The largest $\gamma$
factors are observed in ultra high-energy cosmic rays, reaching $\gamma\sim 10^{11}$ \cite{Olinto:2000sa}. In
this case, the exponential in Eq.\ (\ref{dEdx}) can be set to unity for all $D>6$. Using the above constraints,
we obtain
\begin{equation}
\frac{d\ln{E}}{dx}\gtrsim 10^{19-120/(D-4)} \left(\frac{L}{l}\right)^{D}\left(\frac{E}{10^{20}~{\rm
eV}}\right)~{\rm Mpc}^{-1}\,.
\label{inUHECR}
\end{equation}
Equation (\ref{inUHECR}) imposes stringent limits on the smoothness of the brane. Upper bounds on $L/l$ range
from $\sim 10^3$ for $D=7$ to $\lesssim 1$ for $D\ge 9$. Therefore, GDR implies that the brane must be smooth
on scales of order of the size of the large extra dimensions, i.e., on scales much larger than the Planck
length. This fact has important consequences for the problem of initial conditions in models of brane
cosmology. (See, e.g, Ref.\ \cite{Khoury:2001wf}.)
%%%%%%%%%%%%%%%%%%%%%%%%%%%%%%%%%%%%%%%%%%%%%%%%%%%%%%%%%%%%%%%%%%%%%%%%%%%%%%%%%%%%%%%%%%%%%%%%%%%%%%%%%%%%%%%
\section{Discussion}
\label{discussion}
%%%%%%%%%%%%%%%%%%%%%%%%%%%%%%%%%%%%%%%%%%%%%%%%%%%%%%%%%%%%%%%%%%%%%%%%%%%%%%%%%%%%%%%%%%%%%%%%%%%%%%%%%%%%%%
In higher-dimensional models of our universe, particles radiate gravitational waves due to the structure of the
extra dimensions, specifically the presence of an inhomogeneous hidden brane. We have called this phenomenon
``Gravitational Diffraction Radiation" because of its similarity with the electromagnetic diffraction radiation
which is emitted by a charge moving near a metallic grating. GDR is a general phenomenon of all
higher-dimensional braneworld scenarios. It applies to particles propagating both on the brane and in the bulk.
As an example of its physical effects, we have discussed the GDR emitted by a particle on the visible brane and
derived lower bounds on the scale of brane perturbations.

The above results can be extended in several directions. For instance, this paper considers only inhomogeneities
on the hidden brane. The visible brane is also inhomogeneous. Therefore, free motion on the visible brane is not
uniform from the five-dimensional perspective. This effect could be comparable to the GDR effect studied here. The
derivation of GDR presented above is purely classical. Although the scale of brane inhomogeneities suggests that
the system can be treated as classical, the importance of quantum effects should be assessed. It would also be
interesting to consider GDR emitted by photons or bulk particles. In the latter case, GDR energy loss would
manifest itself as a background of stochastic gravitational waves on the visible brane.

GDR is originated by the lumpiness of the spacetime at small scales. Thus GDR effects may show up in a variety
of physical situations besides braneworld models. Emission of radiation by free particles in a cosmic string
background was discussed in Ref.\ \cite{aliev}. GDR could also play an important role in the very early
universe or near the horizon of black holes. Direct signatures are unlikely to be observed in these cases.
However, GDR emission could lead to interesting predictions at theoretical level and possibly indirect
observable effects.
%%%%%%%%%%%%%%%%%%%%%%%%%%%%%%%%%%%%%%%%%%%%%%%%%%%%%%%%%%%%%%%%%%%%%%%%%%%%%%%%%%%%%%%%%%%%%%%%%%%%%%%%%%%%%%
\section*{Acknowledgements}
%%%%%%%%%%%%%%%%%%%%%%%%%%%%%%%%%%%%%%%%%%%%%%%%%%%%%%%%%%%%%%%%%%%%%%%%%%%%%%%%%%%%%%%%%%%%%%%%%%%%%%%%%%%%%%
We thank Alikram Aliev, Luca Bombelli and Roy Maartens for discussions and useful suggestions. VC acknowledges
financial support from Funda\c c\~ao Calouste Gulbenkian through the Programa Gulbenkian de Est\'{\i}mulo \`a
Investiga\c c\~ao Cient\'{\i}fica.
%%%%%%%%%%%%%%%%%%%%%%%%%%%%%%%%%%%%%%%%%%%%%%%%%%%%%%%%%%%%%%%%%%%%%%%%%%%%%%%%%%%%%%%%%%%%%%%%%%%%%%%%%%%%%

\end{document}